\documentclass[aps,prd,nofootinbib,preprint]{revtex4}
\usepackage{graphicx}

\begin{document}

\title{Constraints on the Light Pseudoscalar Meson Distribution Amplitudes from Their Meson-Photon Transition Form Factors}

\date{\today}

\author{Xing-Gang Wu}
\email[email: ]{wuxg@cqu.edu.cn}
\affiliation{Department of Physics, Chongqing University, Chongqing 401331, P.R. China}
\author{Tao Huang}
\email[email: ]{huangtao@ihep.ac.cn}
\affiliation{Institute of High Energy Physics and Theoretical Physics Enter for Science Facilities, Chinese Academy of Sciences, Beijing 100049, P.R. China}

\begin{abstract}

The meson-photon transition form factors $\gamma\gamma^*\to P$ ($P$ stands for $\pi$, $\eta$ and $\eta'$) provide strong constraints on the distribution amplitudes of the pseudoscalar mesons. In this paper, these transition form factors are calculated under the light-cone perturbative QCD approach, in which both the valence and non-valence quarks' contributions have been taken into consideration. To be consistent, an unified wavefunction model is adopted to analyze these form factors. It is shown that with proper charm component $f^{c}_{\eta'}\sim -30$ MeV and a moderate DA with $B\sim 0.30$, the experimental data on $Q^{2}F_{\eta\gamma}(Q^2)$ and $Q^{2}F_{\eta'\gamma}(Q^2)$ in whole $Q^2$ region can be explained simultaneously. Further more, a detailed discussion on the form factors' uncertainties caused by the constituent quark masses $m_q$ and $m_s$, the parameter $B$, the mixing angle $\phi$ and $f_{\eta'}^c$ are presented. It is found that by adjusting these parameters within their reasonable regions, one can improve the form factor to a certain degree but can not solve the puzzle for $Q^{2}F_{\pi\gamma}(Q^2)$, especially to explain the behavior of $\pi-\gamma$ form factor within the whole $Q^2$ region consistently. We hope further experimental data on these form factors in the large $Q^2$ region can clarify the present situation.

\begin{description}

\item[PACS numbers] 12.38.-t, 12.38.Bx, 14.40.Be

\end{description}

\end{abstract}

\maketitle

\section{Introduction}

The distribution amplitude (DA) is the key factor for exclusive processes. Usually, the DAs of the light pseudoscalar mesons can be expanded in Gegenbauer polynomials, and their corresponding Gegenbauer moments have been studied by various groups, cf.
Refs.\cite{a2a40,a2a41,a2a42,a2a43,a2a44,a2a45,a2lattice1,a2lattice2,a2lattice3}.
However, there is no definite conclusion on whether the shape of the DA is in an asymptotic form \cite{lb} or in a more broad form \cite{cz}.

The light pseudoscalar meson-photon transition form factor $F_{P\gamma}(Q^2)$ that describes the effect of the strong interaction on $\gamma\gamma^*\to P$ transition ($P$ stands for $\pi$, $\eta$ and $\eta'$) provides a good platform for studying the leading-twist DA, since it contains only one bound state and the power suppressed light meson's higher helicity and higher twist structures usually give negligible contributions. One can extract useful information on the leading-twist DA by comparing the theoretical with the measured data on $F_{P\gamma}(Q^2)$.

Based on CELLO, CLEO and the BABAR data on $\gamma\gamma^*\to \pi$ \cite{cello,cleo,babar}, many people have discussed the properties of pion DA
\cite{flatda,wh,brod1,brod2,mikha1,klusek,bakulev,zh,flatda2,new1,new2,new3,new4,new15}. A CZ-like DA or even a flat DA \cite{flatda} can explain the large $Q^2$ behavior shown by the BABAR data \cite{babar}, however the theoretical estimation on the form factor with a CZ-like or flat DA shall always be lower than the experimental results in small $Q^2$ region \cite{flatda2}. While by taking the non-valence quark parts into consideration, it is found that by setting the second pion moment $a^{\pi}_2(\mu^2_0)$ around $0.35$, one can explain the behavior in small $Q^2$ region well, however a somewhat large discrepancy emerges in high $Q^2$ region if the BABAR data is confirmed \cite{wh}. A reasonable theoretical estimation on the form factor should explain the measured form factor's behaviors in both the lower and higher $Q^2$-regions consistently. And one should find a way to compare the experiment results on the form factor to determine which DA shape is more suitable.

More over, the experimental data on $\gamma\gamma^*\to \eta$ or $\eta'$ transition
\cite{cello,cleo,l3,tpc,kloe,kloe1,babarold,pluto,bes}, especially, the new BABAR results within the region of $[4, 40]$ GeV$^2$ \cite{babareta}, can provide further constraints on the pseudoscalar meson's DA \cite{whtheta,new5,new6,new7,new8,new9,new10}. Then, we shall have three pseudoscalar meson-photon transition form factors $F_{P\gamma}(Q^2)$ to constrain the light meson's wavefunction (WF) / DA parameters. Because of $\eta$ and $\eta'$ mixing, their condition is somewhat more difficult than the pionic case. Two mixing schemes are adopted in the literature, even though their mixing parameters can be related through the correlation given by Refs.\cite{oct1,oct2}, there are differences in dealing with certain processes. One mixing scheme is based on the flavor singlet $\eta_1$ and octet $\eta_8$, under which, one usually introduces two mixing angles $\theta_1$ and $\theta_8$ and adopts the same DA for $\eta_1$ and $\eta_8$ \cite{new5,new6,new7,new8,new9,new10}. In the present paper, we adopt the simpler quark-flavor mixing scheme \cite{oct1}, which is based on the quark-flavor basis $\eta_q$ and $\eta_s$ and only one mixing angle $\phi$ is introduced. Since $\eta_q$ and $\eta_s$ have similar structure as that of pion, it is natural to adopt the same WF model for $\pi$, $\eta_q$ and $\eta_s$. For the purpose, we adopt the WF model raised by Ref.\cite{wh} to do our discussion, since by setting the parameter $B$ properly, we can obtain different DA behavior from AS-like to CZ-like naturally, and then to determine which one is more suitable for simultaneously explaining the experimental data of these form factors.

The paper are organized as follows. In Sec.II, we outline our calculation techniques for the transition form factors $F_{P\gamma}(Q^2)$, where the mixing scheme for $\eta$ and $\eta'$, an uniform WF model for the mesons, and the analytic formulas for deriving $F_{P\gamma}(Q^2)$ are presented. In Sec.III, we present the numerical results, and discuss the uncertainty sources. The final section is reserved for a summary.

\section{Basic formulas for the form factors $F_{P\gamma}(Q^2)$}

In this section, we present necessary formulas for the transition form factors $F_{P\gamma}(Q^2)$. First, we define the physical meson states $\eta$ and $\eta'$ under the quark-flavor basis, then we give the uniform WF model for the mentioned pseudoscalar mesons, and finally, we present the form factor with the valence quark contribution calculated up to next-to-leading order (NLO) together with an estimation of the non-valence quark state's contributions.

\subsection{$\eta$ and $\eta'$ defined under the quark-flavor basis}

The physical meson states $\eta$ and $\eta'$ are related to the orthogonal states $\eta_q$ and $\eta_s$ through an orthogonal transformation
\begin{eqnarray}
\left( \matrix{|\eta\phantom{{}'}\rangle \cr |\eta'\rangle } \right)
&=& U(\phi) \, \left(\matrix{|\eta_q\rangle \cr|\eta_s\rangle }
\right) \; ,\;\;\;  U(\phi) = \left(\matrix{\cos\phi & -\sin\phi \cr
\sin\phi & \phantom{-}\cos\phi} \right) \ , \label{mixangle}
\end{eqnarray}
where $\phi$ is the mixing angle. Here, we adopt a single mixing angle scheme that attributes $SU_{f}(3)$ breaking to the Okubo-Zweig-Iizuka violating contribution \cite{oct1}. In the quark-flavor basis, the two orthogonal states $|\eta_q \rangle$ and $|\eta_s \rangle$ are defined in a Fock state description, $|\eta_q \rangle = \Psi_{\eta_q} \, \frac{|u\bar u + d\bar d\rangle}{\sqrt{2}}$ and $|\eta_s \rangle = \Psi_{\eta_s} |s\bar s \rangle$, where $\Psi_{\eta_i}$ ($i=q$, $s$) denote the light-cone WFs of the corresponding parton states. Under such scheme, the decay constants in the quark-flavor basis simply follow the pattern of state mixing, i.e.
\begin{eqnarray}
\left(\begin{array}{cc}f_\eta^q & f_\eta^s \cr f_{\eta'}^q &
f_{\eta'}^s\end{array} \right) &=& U(\phi) \, {\rm diag}[f_{\eta_q},f_{\eta_s}] ,
\label{decayconstant}
\end{eqnarray}
where $f_{\eta_i} = (2\sqrt{3}) \int_{k_\perp^2\leq \mu_0^2} \frac{dx d^2k_\perp}{16 \pi^3} \Psi_{\eta_i}(x,k_\perp)$ and the factorization scale $\mu_0\sim{\cal O} (1\;{\rm GeV})$.

One can obtain the correlation between $f_{\eta_q}/f_{\eta_s}$ and $\phi$ from the two-photon decay of $\eta$ and $\eta'$, i.e. $\eta/\eta'\to\gamma\gamma$, which shows \cite{whtheta}
\begin{eqnarray}\label{decayfq}
f_{\eta_q} &=& \frac{c_q\alpha}{8\pi^{3/2}}\left[\sqrt{\frac{\Gamma_{\eta\to\gamma\gamma}} {M_\eta^3}}\cos\phi+\sqrt{\frac{\Gamma_{\eta'\to\gamma\gamma}}{M_{\eta'}^3}}
\sin\phi\right]^{-1}
\end{eqnarray}
and
\begin{eqnarray}
f_{\eta_s} &=& \frac{c_s\alpha}{8\pi^{3/2}}\left[\sqrt{
\frac{\Gamma_{\eta'\to\gamma\gamma}}{M_{\eta'}^3}}\cos\phi-\sqrt{
\frac{\Gamma_{\eta\to\gamma\gamma}}{M_\eta^3}}\sin\phi\right]^{-1}
,\label{decayfs}
\end{eqnarray}
where $c_s=\sqrt{2}/3$, $c_q=5/3$ and $\alpha=1/137$. Since the power suppressed higher twist and higher helicity components give negligible contributions to the two-photon decay of $\eta$ and $\eta'$, the correlations (\ref{decayfq},\ref{decayfs}) shall provide strong constraint on $f_{\eta_q}$, $f_{\eta_s}$ and $\phi$.

\subsection{wavefunction of the light pseudoscalar meson}

As for the light pseudoscalar meson ($P$), its light-cone WF can be written as
\begin{equation}\label{wave}
\Psi_{P}(x,{\bf k}_{\perp})=\sum_{\lambda_{1}\lambda_{2}} \chi^{\lambda_{1}\lambda_{2}}(x,m_{i},{\bf k}_{\perp}) \Psi^{R}_{P}(x,m_i,{\bf k}_{\perp}) ,
\end{equation}
where $i$ stands for the light quark $q$ or $s$, $\lambda_1$ and $\lambda_2$ are helicities of the two constituent quarks. $\chi^{\lambda_{1}\lambda_{2}}(x,{\bf k}_{\perp})$ stands for the spin-space WF coming from the Wigner-Melosh rotation. $\Psi^{R}_{q\bar{q}}(x,m_{i},{\bf k}_{\perp})$ stands for the spatial WF, which can be factorized as \cite{wh}
\begin{widetext}
\begin{equation}\label{wave1}
\Psi^{R}_{P}(x,m_i,{\bf k}_{\perp})=A\varphi_P(x) \exp\left[-\frac{{\bf k}_{\perp}^2 +m_i^2} {8{\beta_i}^2x(1-x)}\right].
\end{equation}
\end{widetext}
The $x$-dependent $\varphi_P(x)$ can be expanded in Gegenbauer polynomials, and by keeping its first two terms, we obtain
\begin{widetext}
\begin{equation}
\Psi^{R}_{P}(x,m_{i},{\bf k}_{\perp})=A\left(1+B\times C^{3/2}_2(2x-1)\right) \exp\left[-\frac{{\bf k}_{\perp}^2 +m_i^2}{8{\beta}^2x(1-x)}\right] .
\end{equation}
\end{widetext}
As for the parameters $B$ and $\beta$, $B$ determines the broadness of the WF in the longitudinal direction, while $\beta$ determines the WF's transverse behavior.

The DA $\phi_P(x)$ can be obtained from $\Psi_P(x,\mathbf{k_{\perp}})$ by integrating over the transverse momentum, $\phi_P(x)=\int_{|\mathbf{k}_\perp|<\mu_0}
\frac{d^2\mathbf{k_\perp}}{16\pi^3} \left(\frac{2\sqrt{3}}{f_P}\right) \Psi_P(x,\mathbf{k_\perp})$, and we obtain
\begin{eqnarray}
\phi_P(x,\mu_0^2) &=& \frac{\sqrt{3}A m \beta}{2\sqrt{2}\pi^{3/2}f_P} \sqrt{x(1-x)} \left(1+B\times C^{3/2}_2(2x-1)\right) \cdot \nonumber\\
&& \left( \mathrm{Erf} \left[\sqrt{\frac{m^2+\mu_0^2}{8\beta^2 x(1-x)}}\right]- \mathrm{Erf}\left[\sqrt{\frac{m^2}{8\beta^2 x(1-x)}}\right] \right),  \label{ourphi}
\end{eqnarray}
where $\mathrm{Erf}(x)=\frac{2}{\sqrt{\pi}} \int_0^x e^{-t^2}dt$. $\phi_P(x,\mu^2_0)$ can be expanded in conventional Gegenbauer polynomials, whose moments $a_n(\mu^2_0)=\frac{\int_0^1 dx \phi_{P}(x,\mu^2_0)C^{3/2}_n(2x-1)} {\int_0^1 dx 6x(1-x) [C^{3/2}_n(2x-1)]^2}$. Numerically, it is found that its second Gegenbauer moment $a_2(\mu^2_0)$ is close to $B$, i.e. the DA's behavior is dominated by $B$. Moreover, when $B\simeq0.00$, its DA is asymptotic-like; and when $B\simeq0.60$, its DA is CZ-like. This shows $\phi_P(x,\mu^2_0)$ can mimic the DA behavior from asymptotic-like to CZ-like naturally. Then, by comparing the estimations for $B\in[0.00,0.60]$ with the experimental data on various processes, one may decide which is the right DA behavior possessed by the light pseudoscalar mesons. Here, we do not discuss the flat DA, since it is hard to explain the meson-photon transition form factor's behavior around $Q^2\sim0$ and shall meet even more serious end-point problem than the CZ-like DA at $x\sim 0,\;1$ \cite{wh}.

\subsection{Pseudoscalar-photon transition form factors}

The pseudoscalar meson-photon transition form factors can be divided into two parts
\begin{equation}\label{fft}
F_{P \gamma}(Q^2)=F^{(V)}_{P\gamma}(Q^2)+ F^{(NV)}_{P\gamma}(Q^2) ,
\end{equation}
where $F^{(V)}_{P\gamma}(Q^2)$ is the valence quark part, $F^{(NV)}_{P\gamma}(Q^2)$ stands for the non-valence quark part that is related to the higher Fock state of the pseudoscalar meson.

Under the light-cone perturbative QCD approach \cite{lb}, and by keeping the $k_\bot$-corrections in both the hard-scattering amplitude and the WF, $F_{\pi \gamma}(Q^2)$ has been calculated up to NLO \cite{bhl,huangwu,huang1,rady1,hnli1,hnli2}. For pseudoscalar meson-photon transition form factors up to NLO, we have \cite{wh},
\begin{eqnarray}
F^{(V)}_{P \gamma}(Q^2)&=& \frac{\sqrt{3}e_P}{4\pi^2}\int_0^1\int_0^{x^2 Q^2}\frac{dx}{x Q^2}\left[1-\frac{C_F \alpha_s(Q^2)}{4\pi}\left(\ln\frac{Q^2}{xQ^2+k_\perp^2} +2\ln{x}+3- \frac{\pi^2}{3} \right)\right] \cdot \nonumber\\
& & \Psi_{P}(x,k_\perp^2) d k^2_\perp , \label{ffv}
\end{eqnarray}
where $[dx]=dxdx'\delta(1-x-x')$, $C_F=4/3$ and $k_\perp=|\mathbf{k}_\perp|$. And $e_P$ relates to the electric charge of the constituent quarks, $e_{\pi}=1/3$, $e_{\eta_q}=5/9$ and $e_{\eta_s}=\sqrt{2}/9$.

As for $F^{(NV)}_{P \gamma}(Q^2)$, we adopt the model suggested by Refs.\cite{huangwu}, which is constructed based on the form factor's limiting behavior at both $Q^2\to 0$ and $Q^2\to\infty$, i.e.,
\begin{equation}\label{ffnv}
F^{(NV)}_{P \gamma}(Q^2)=\frac{\alpha}{(1+Q^2/\kappa^2)^2} ,
\end{equation}
where $\alpha=\frac{1}{2}F_{P\gamma}(0)$, $\kappa=\sqrt{-\frac{F_{P\gamma}(0)} {F^{(NV)'}_{P\gamma}(Q^2)|_{Q^2\to 0}}}$ with the first derivative of $F^{(NV)}_{P \gamma}(Q^2)$ over $Q^2$ in the limit $Q^2\to 0$ takes the form
\begin{displaymath}
F^{(NV)'}_{P\gamma}(Q^2)|_{Q^2\to 0}=\frac{\sqrt{3}e_{P}}{8\pi^2}\left[\frac{\partial}{\partial
Q^2}\int_0^1\int_{0}^{x^2 Q^2}\left(\frac{\Psi_{P}(x,k_\perp^2)}{x^2 Q^2}\right) dx dk_\perp^2 \right]_{Q^2\to 0}.
\end{displaymath}

The same phenomenological model for $F^{(NV)}_{P \gamma}(Q^2)$ has also been adopted by Ref.\cite{brod1}, where instead a fixed input parameter $\Lambda\sim1.1$ GeV is introduced to replace the parameter $\kappa$. Since the octet-singlet mixing scheme \cite{oct1,oct2} is adopted by Ref.\cite{brod1}, it is reasonable to take the same $\Lambda$ for $\pi$, $\eta_8$ and $\eta_0$. Under the present adopted quark-flavor mixing scheme, we shall numerically obtain $\kappa\sim 1.1-1.2$ GeV for pion and $\eta_q$, and $\kappa\sim 1.5-1.6$ GeV for $\eta_s$, where different $\kappa$ is rightly caused by $SU_f(3)$-breaking effect.

Moreover, under the quark-flavor mixing scheme, $\eta-\gamma$ and $\eta'-\gamma$ transition form factors are related with $F_{\eta_{q}\gamma}(Q^2)$ and $F_{\eta_{s}\gamma}(Q^2)$ through the following equations
\begin{eqnarray}
F_{\eta\gamma}(Q^2)&=&F_{\eta_q\gamma}(Q^2)\cos\phi -F_{\eta_s\gamma}(Q^2)\sin\phi \label{ffeta}
\end{eqnarray}
and
\begin{eqnarray}
F_{\eta'\gamma}(Q^2)&=&F_{\eta_q\gamma}(Q^2)\sin\phi +F_{\eta_s\gamma}(Q^2)\cos\phi .\label{ffetap}
\end{eqnarray}

\section{Numerical results and discussions}

\subsection{Input parameters}

Two photon decay widths of $\eta$ and $\eta'$, and their masses can be found in PDG \cite{pdg}
\begin{eqnarray}
&&\Gamma_{\eta\to\gamma\gamma}=0.510\pm0.026 \;{\rm KeV},\;\;
M_{\eta}=547.853\pm0.024 \;{\rm MeV}, \nonumber\\
&&\Gamma_{\eta'\to\gamma\gamma}=4.28\pm0.19\;{\rm KeV},\;\;
M_{\eta'}=957.78\pm0.06 \;{\rm MeV}, \nonumber
\end{eqnarray}
and $f_\pi=92.4\pm0.25$ MeV.

A weighted average of the experimental values shown in Ref.\cite{feldmann}, together with two experimental values $\phi=38.8^{\circ}\pm 1.2^{\circ}$ \cite{bes} and $\phi=41.2^{\circ}\pm1.1^{\circ}$ \cite{kroll}, yields $\bar\phi={39.5^{\circ}} \pm {0.5^{\circ}}$. Then, with the help of Eqs.(\ref{decayfq},\ref{decayfs}), we obtain
\begin{equation}
f_{\eta_q}/f_{\pi}=1.07\pm0.03
\end{equation}
and
\begin{equation}
f_{\eta_s}/f_{\pi}=1.44\pm 0.08 ,
\end{equation}
which are consistent with the ``phenomenological" values, $f_{\eta_q}\simeq f_{\pi}$ and $f_{\eta_s}\simeq 1.36 f_{\pi}$ \cite{feldmann}.

\begin{table}
\caption{Typical WF parameters for $m_q=0.30$ GeV and $m_s=0.45$ GeV, where $\phi=39.5^{\circ}$ is adopted.}
\begin{center}
\begin{tabular}{|c||c|c|c|c|}
\hline\hline ~~~$B|m$~~~ & ~$\beta_\pi({\rm GeV})$~ & ~$A_{\pi}({\rm GeV}^{-1})$~ & ~$A_{q}({\rm GeV}^{-1})$~ & ~$A_{s}({\rm GeV}^{-1})$~\\
\hline
~$0.00|m_q$~ & ~$0.586$~& ~$25.06$~ & ~$26.81$~ & ~/~\\
\hline
~$0.30|m_q$~ & ~$0.668$~& ~$20.26$~ & ~$21.67$~ & ~/~ \\
\hline
~$0.60|m_q$~ & ~$0.745$~& ~$16.62$~ & ~$17.78$~ & ~/~\\
\hline\hline
~$0.00|m_s$~ & ~$0.464$~& ~$42.23$~ & ~/~ & ~$60.58$~\\
\hline
~$0.30|m_s$~ & ~$0.504$~& ~$36.97$~ & ~/~ & ~$49.58$~ \\
\hline
~$0.60|m_s$~ & ~$0.552$~& ~$31.24$~ & ~/~ & ~$40.72$~\\
\hline\hline
\end{tabular}
\label{tab1}
\end{center}
\end{table}

As for pion WF, its parameters can be determined by its normalization condition and the constraint from $\pi_{0}\to\gamma\gamma$. A detailed determination on pion WF parameters can be found in Ref.\cite{wh}, where $\beta_\pi$ for specified $B$ and quark mass is determined by
\begin{equation}
\frac{\int^1_0 dx \int_{|\mathbf{k}_\perp|^2<\mu_0^2}
\frac{d^{2}{\bf k}_{\perp}}{16\pi^3}\Psi_{\pi}(x,{\bf
k}_{\perp})}{\int^1_0 dx \Psi_{\pi}(x,{\bf k}_{\perp}=0)} =\frac{f^2_{\pi}}{6} .
\end{equation}
Because $\eta_q$ and $\eta_s$ have similar behaviors as that of $\pi$, for clarity, we take $\beta_{q}=\beta_{\pi}|_{m_q}$ and $\beta_{s}=\beta_{\pi}|_{m_s}$. Under the condition of $B=0.00$, $0.30$ and $0.60$, typical parameters for the DAs of $\pi$, $\eta_q$ and $\eta_s$ are presented in TAB.(\ref{tab1}), where the mixing angle $\phi$ is fixed to be $39.5^{\circ}$.

\begin{figure}
\centering
\includegraphics[width=0.45\textwidth]{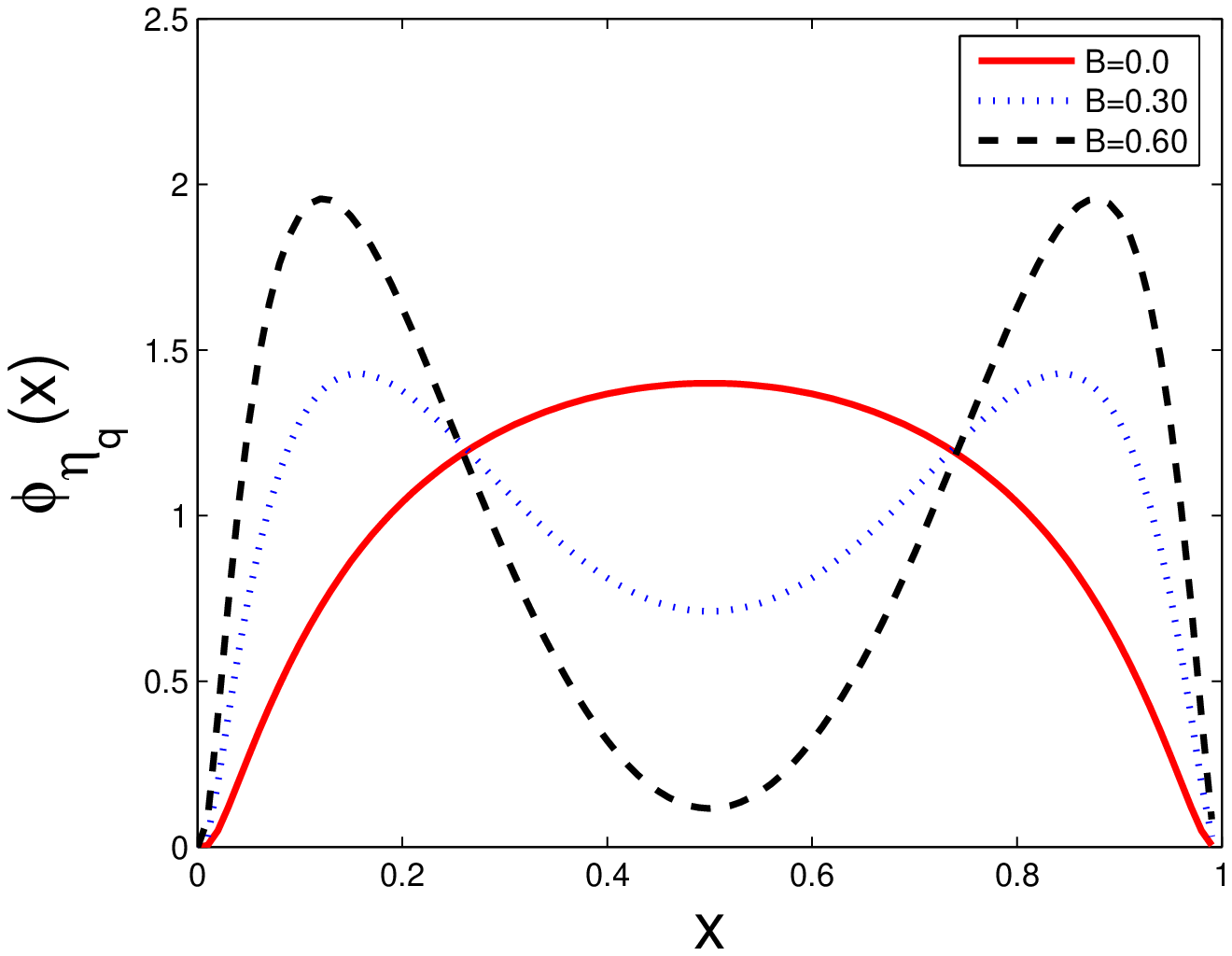}
\includegraphics[width=0.45\textwidth]{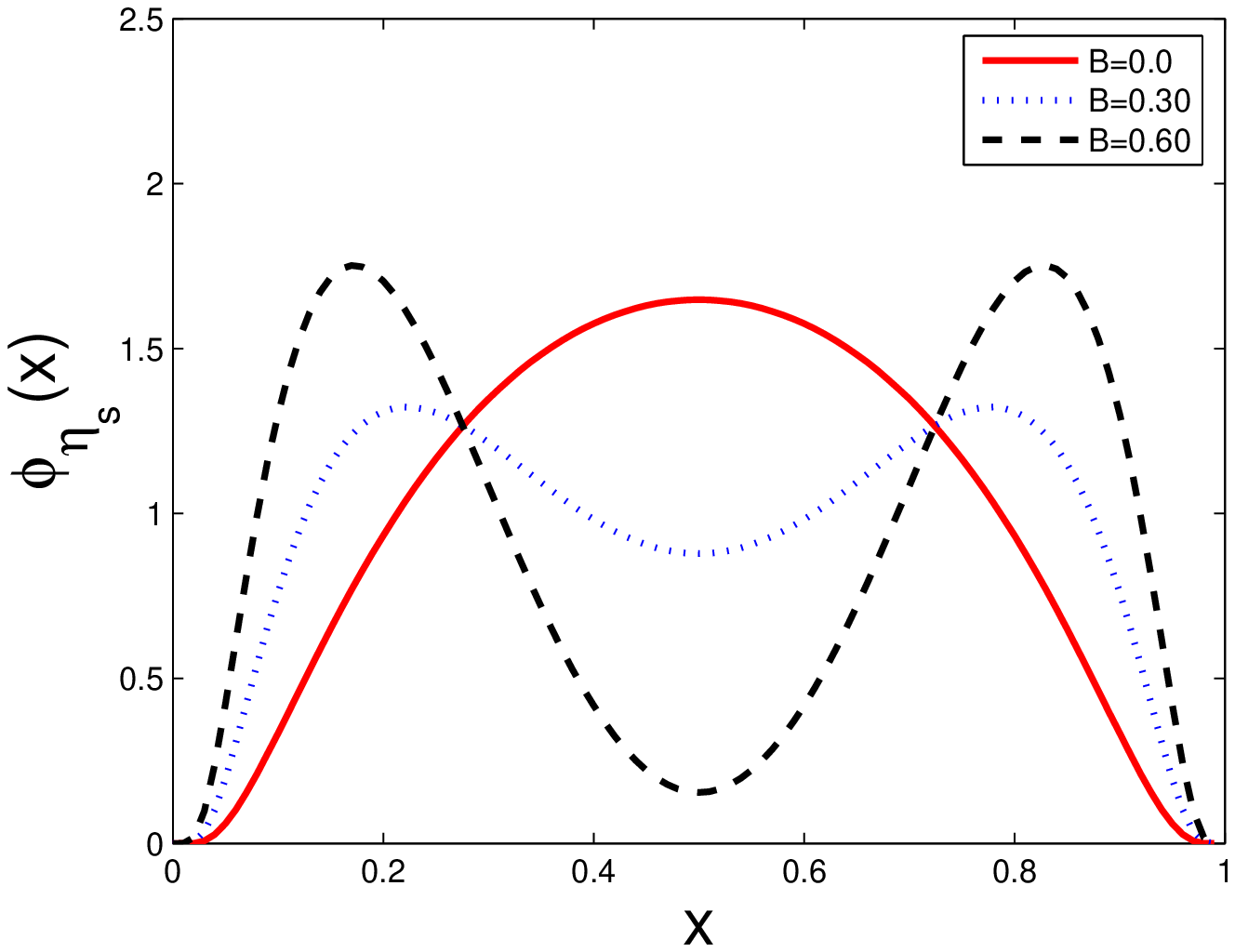}
\caption{DA model (\ref{ourphi}) for $\phi_{\eta_q}(x)$ (Left) and $\phi_{\eta_s}(x)$ (Right) with $\mu_0=1$ GeV, where $B=0.00$, $0.30$ and $0.60$ respectively. } \label{phi}
\end{figure}

It is noted that by varying $B$ within the region of $\sim [0.00,0.60]$, the DAs shall vary from asymptotic-like to CZ-like form. To show this point more clearly, we draw $\phi_{\eta_q}(x)$ and $\phi_{\eta_s}(x)$ in Fig.(\ref{phi}), where $B=0.00$, $0.30$ and $0.60$  respectively.

\subsection{Basic numerical results}

Numerically, one may observe that in the large $Q^2$ region, the leading valence Fock-state contribution dominates the form factor $Q^2 F_{P\gamma}(Q^2)$, while the non-valence quark part $Q^2 F^{NV}_{P\gamma}(Q^2)$ is power suppressed and is quite small, so it is usually neglected in the literature. However, $Q^2 F^{NV}_{P\gamma}(Q^2)$ can provide sizable contributions in the low and intermediate energy regions, so one should take it into consideration to make a more sound estimation in the whole energy regions.

As shown above, the parameter $B$ in the unified WF model (\ref{wave}) determines the DA behavior of the light pseudoscalar mesons. Then, by comparing with the experimental data on the pseudoscalar meson-photon transition form factors, it provides us an opportunity to discuss the DA properties in a more consistent way.

\begin{figure}
\centering
\includegraphics[width=0.5\textwidth]{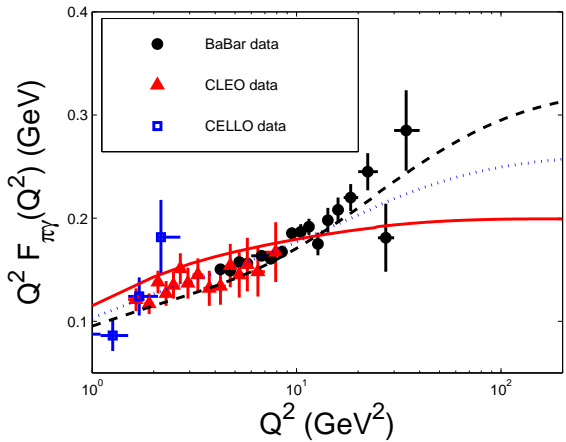}
\caption{$\pi-\gamma$ transition form factor $Q^{2}F_{\pi\gamma}(Q^2)$ with varying $B$. The solid, the dotted and the dashed lines are for $B=0.00$, $0.30$ and $0.60$ respectively. The experimental data are taken from Refs.\cite{cello,cleo,babar}.} \label{pion}
\end{figure}

\begin{figure}
\centering
\includegraphics[width=0.45\textwidth]{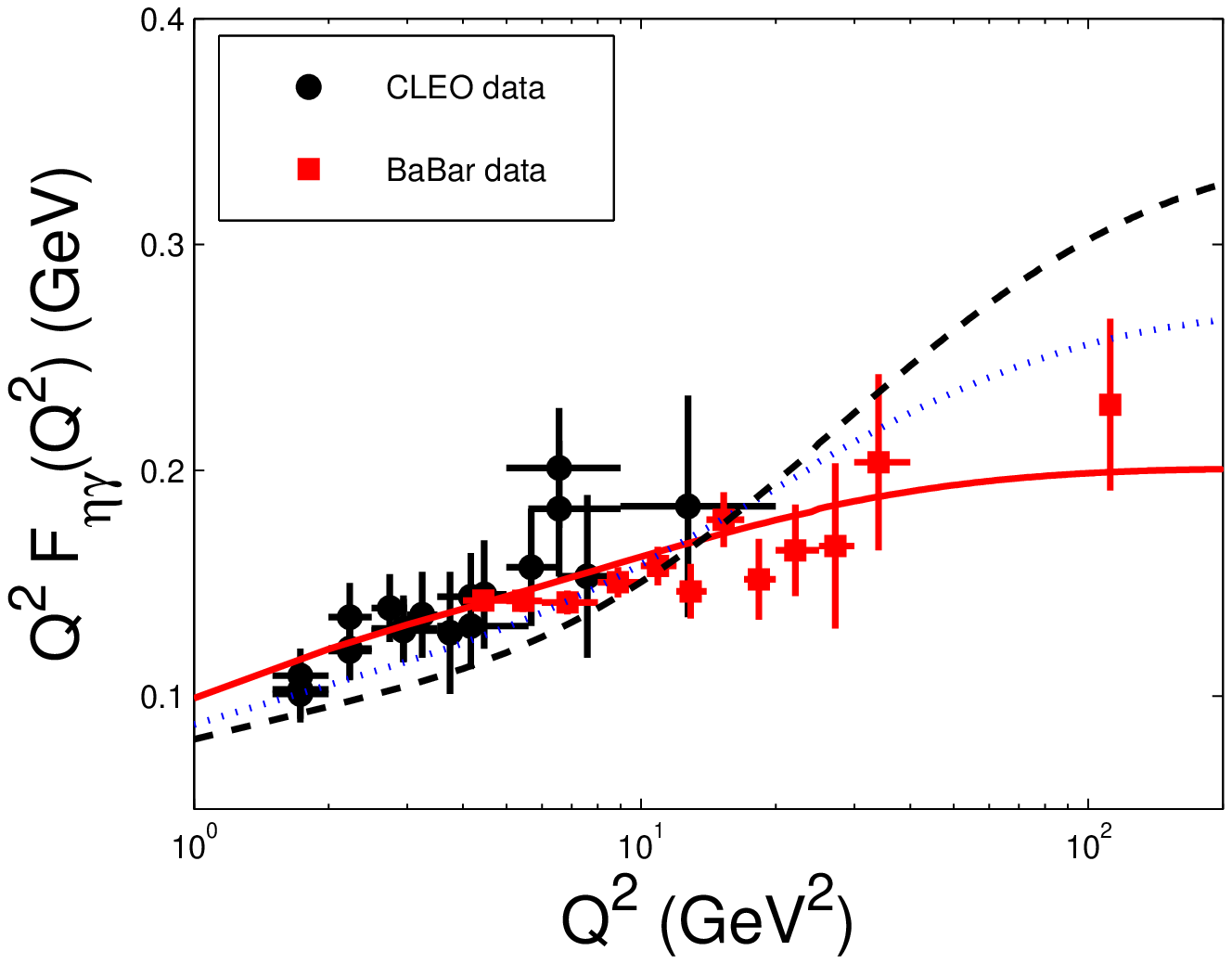}
\includegraphics[width=0.45\textwidth]{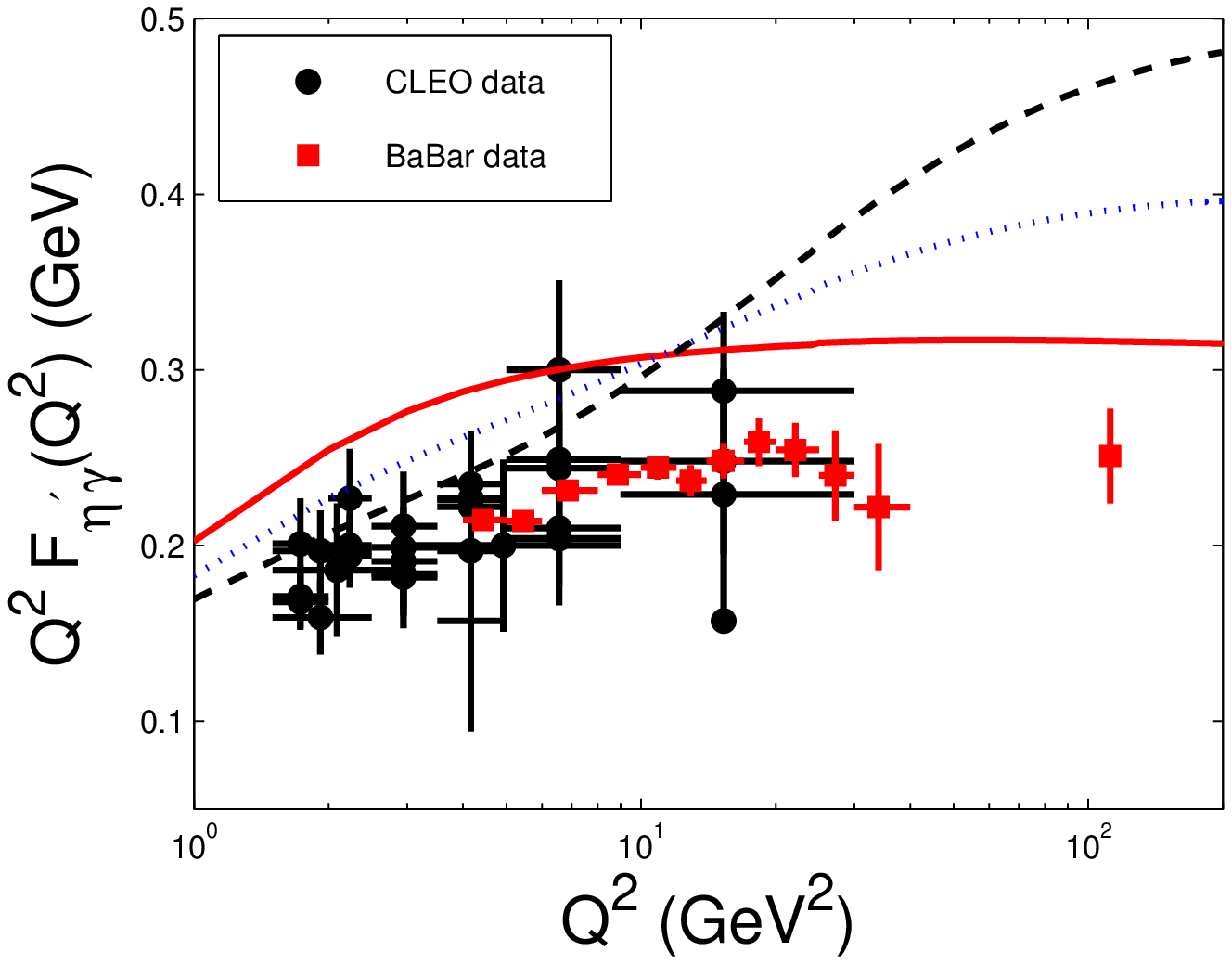}
\caption{$\eta-\gamma$ (Left) and $\eta'-\gamma$ (Right) transition form factors $Q^{2}F_{\eta\gamma}(Q^2)$ and $Q^{2}F_{\eta'\gamma}(Q^2)$. The solid, the dotted and the dashed lines are for $B=0.00$, $0.30$ and $0.60$ respectively. The experimental data are taken from Refs.\cite{cleo,babarold,babareta}. } \label{eta}
\end{figure}

Firstly, we present the meson-photon transition form factors within a wide region of $B\in[0.00,0.60]$ so as to show which DA behavior is more suitable to explain the data, especially the BABAR data \cite{babar}. In doing the numerical calculation, we take all the other input parameters to be their center values, i.e. $m_q=0.30$ GeV, $m_s=0.45$ GeV and $\phi=39.5^{\circ}$. Figs.(\ref{pion}, \ref{eta}) show the pseudoscalar-photon transition form factors $Q^{2}F_{\pi\gamma}(Q^2)$, $Q^{2}F_{\eta\gamma}(Q^2)$ and $Q^{2}F_{\eta'\gamma}(Q^2)$, where $B=0.00$, $0.30$ and $0.60$ respectively. These two figures show that with the increment of $B$, all the three form factors decrease in lower $Q^2$ region but increase in higher $Q^2$ region. Especially, the CZ-like DA ($B=0.60$) leads to the smallest value in the lower $Q^2$ region, while the AS-like one ($B=0.00$) leads to biggest one; and in the higher $Q^2$ region, the condition is vice versa. This causes the present puzzle for explaining the newly obtained BABAR data on $\pi-\gamma$ form factor. The BABAR data shows that in the range of $Q^2\in[4,40]$ GeV$^2$, the pion-photon transition form factor behaves as \cite{babar}, $Q^2 F_{\pi \gamma}(Q^2) = A\left(\frac{Q^2}{10GeV^2}\right)^{\beta}$, where $A=0.182\pm0.002$ and $\beta=0.25\pm0.02$. A CZ-like DA with $B\sim 0.60$ or even flat DA can explain the data well for higher $Q^2$ region \footnote{The form factor with flat DA shall lead to logarithmic growth with $Q^2$, i.e. $Q^2 F_{\pi \gamma}(Q^2)\propto \ln\left(1+Q^{2}/M^{2}\right)$ \cite{flatda2}, which is close to BABAR data. }, however it fails in lower $Q^2$ region. The AS-like DA with $B \sim 0.00$ provides a better understanding for lower $Q^2$ region, however due to the fact that $Q^2 F_{\pi \gamma}(Q^2)$ tends to be a constant value $(2f_{\pi})$, it can not explain the large $Q^2$ behavior. On the one hand, by increasing $B$ from $0$ to a larger value, the estimated large $Q^2$-behavior of $Q^2 F_{\pi \gamma}(Q^2)$ can be improved. On the other hand, the deviation of the lower $Q^2$-behavior also increases with the increment of $B$, so $B$ should not be too large. Moreover, as shown by Fig.(\ref{eta}), $\eta-\gamma$ form factor $Q^{2}F_{\eta\gamma}(Q^2)$ prefers a DA with smaller $B$, i.e. $B \lesssim 0.30$. For the $\eta'-\gamma$ form factor $Q^{2}F_{\eta'\gamma}(Q^2)$, all DAs shall lead to large $Q^2$ behavior well above the experimental data.

\begin{figure}
\centering
\includegraphics[width=0.45\textwidth]{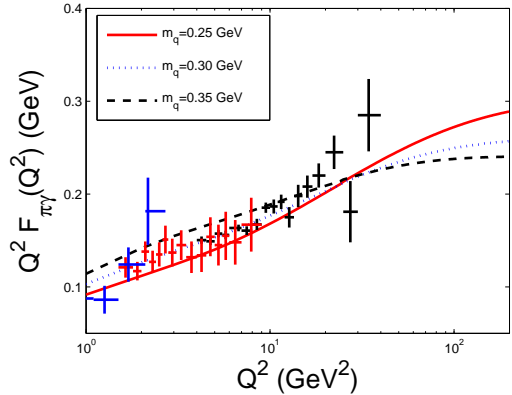}
\caption{$\pi-\gamma$ transition form factor $Q^{2}F_{\pi\gamma}(Q^2)$ with fixed $B=0.3$. The red solid line, the blue dotted line and the black dashed line are for $m_q=0.25$ GeV, $0.30$ GeV and $0.35$ GeV respectively. The experimental data are taken from Refs.\cite{cello,cleo,babar}. } \label{pionmassuncern}
\end{figure}

\begin{figure}
\centering
\includegraphics[width=0.45\textwidth]{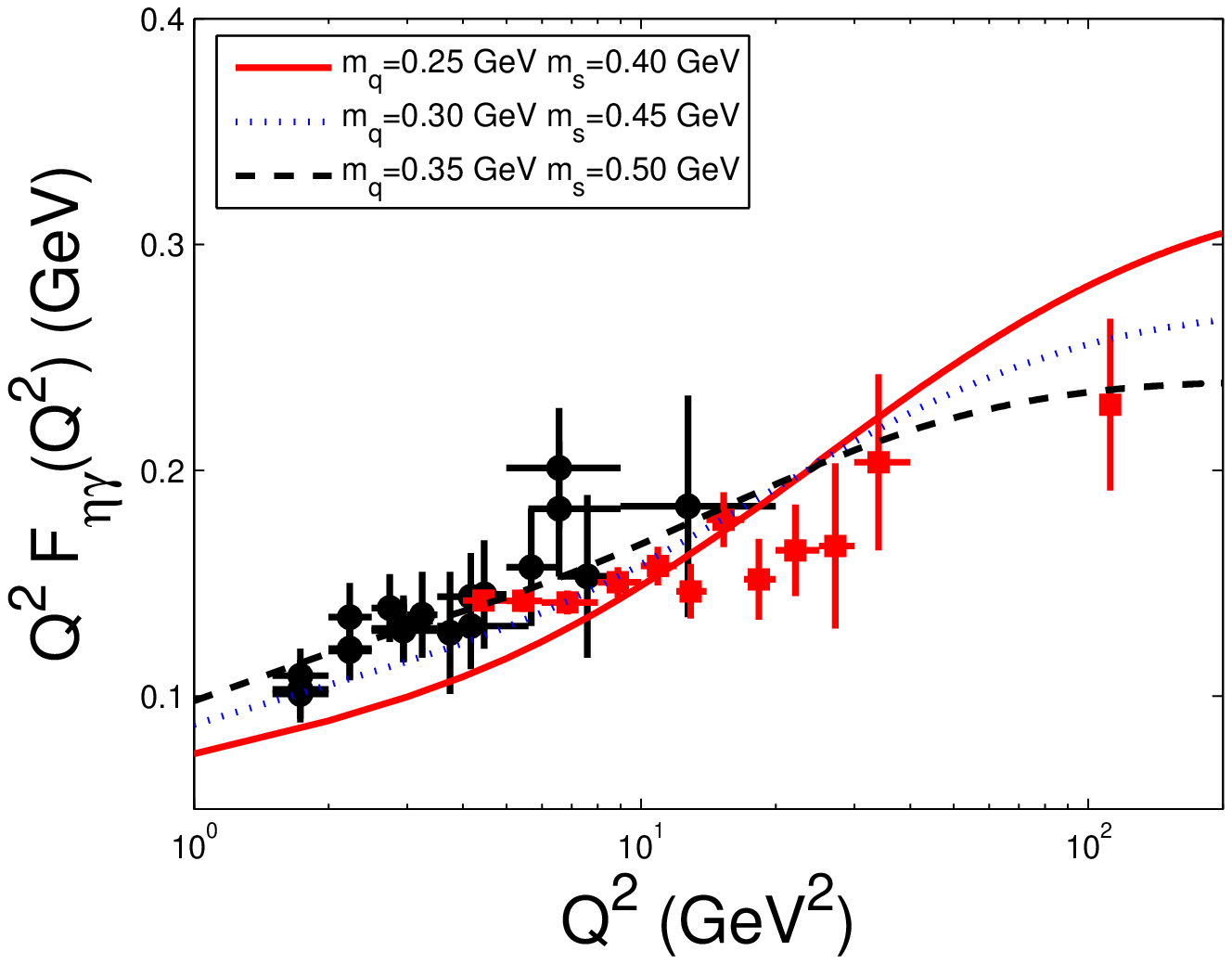}
\includegraphics[width=0.45\textwidth]{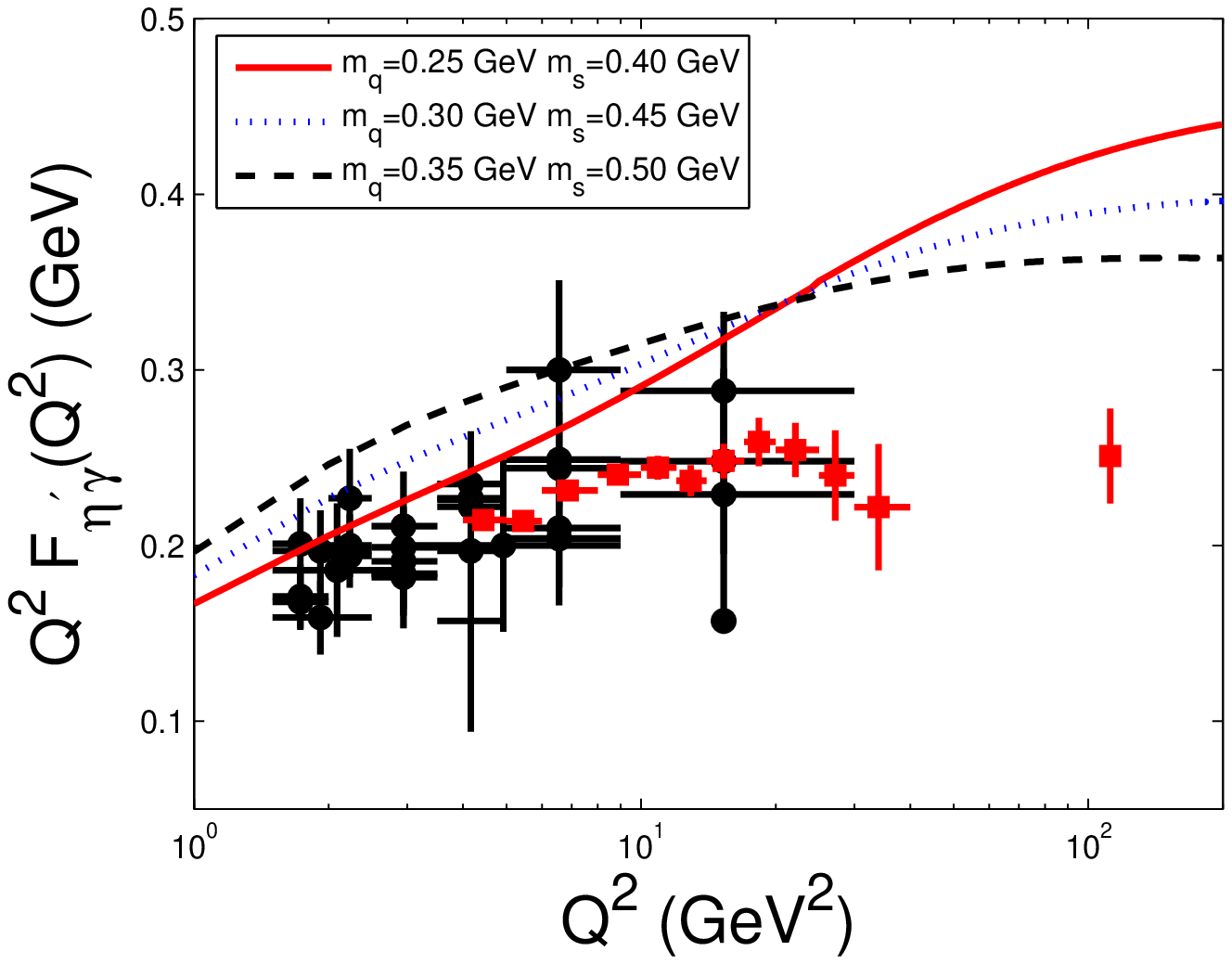}
\caption{$\eta-\gamma$ (Left) and $\eta'-\gamma$ (Right) transition form factors $Q^{2}F_{\eta\gamma}(Q^2)$ and $Q^{2}F_{\eta'\gamma}(Q^2)$ with fixed $B=0.3$ and $\phi=39.5^{\circ}$. The red solid line, the blue dotted line and the black dashed line are for [$m_q=0.25$ GeV and $m_s=0.40$ GeV], [$m_q=0.30$ GeV and $m_s=0.45$ GeV], and [$m_q=0.35$ GeV and $m_s=0.50$ GeV] respectively. The experimental data are taken from Refs.\cite{cleo,babarold,babareta}. } \label{etamassuncern}
\end{figure}

Secondly, we study the uncertainties of the transition form factors caused by $m_q$ and $m_s$. For the purpose, we take $m_q =0.30\pm0.05$ GeV and $m_s =0.45\pm0.05$ GeV, and fix the parameter $B=0.30$ and $\phi=39.5^{\circ}$. The $\pi-\gamma$ form factor for $m_q =0.30\pm0.05$ GeV is presented in Fig.(\ref{pionmassuncern}), and $\eta-\gamma$ and $\eta'-\gamma$ form factors for [$m_q=0.25$ GeV and $m_s=0.40$ GeV], [$m_q=0.30$ GeV and $m_s=0.45$ GeV], and [$m_q=0.35$ GeV and $m_s=0.50$ GeV] are presented in Fig.(\ref{etamassuncern}). It can be found that the form factors change with the constituent quark mass similar to its change with $B$, i.e. with the increment of constituent quark mass, all the three form factors increase in the lower $Q^2$ region and decrease in the higher $Q^2$ region. Naively, one may expect to obtain a larger high $Q^2$ behavior by setting a smaller $m_q$ and $m_s$. Especially, by setting the limiting values of $m_q=0$ and $B=0$, which rightly corresponds to a flat $\varphi_\pi$ as suggested by Ref.\cite{flatda2}, one can obtain the same logarithmic behavior for large $Q^2$ region that is consistent with BABAR data, i.e. $Q^2 F_{\pi \gamma}(Q^2)\propto \ln\left(Q^{2}/2\sigma^{2}\right)$ with $\sigma=\frac{M^2}{2}e^{\gamma_E}$. However, it is found that $m_q$ can not be too small, e.g. it should be larger than $0.22$ GeV, otherwise the probability of leading valence quark state $|q\bar{q}\rangle$ shall be larger that $1$ \cite{hwem}.

\begin{figure}
\centering
\includegraphics[width=0.45\textwidth]{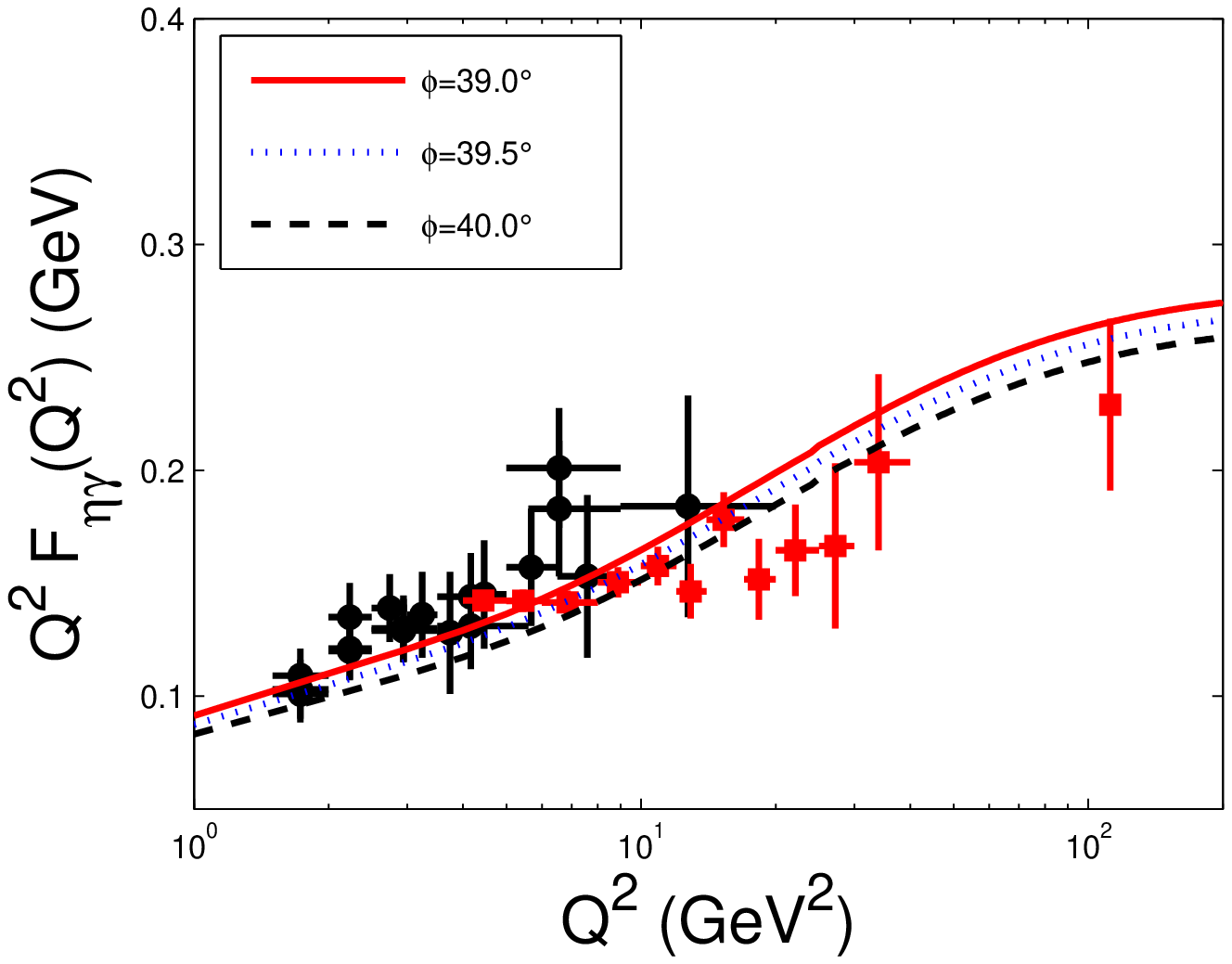}
\includegraphics[width=0.45\textwidth]{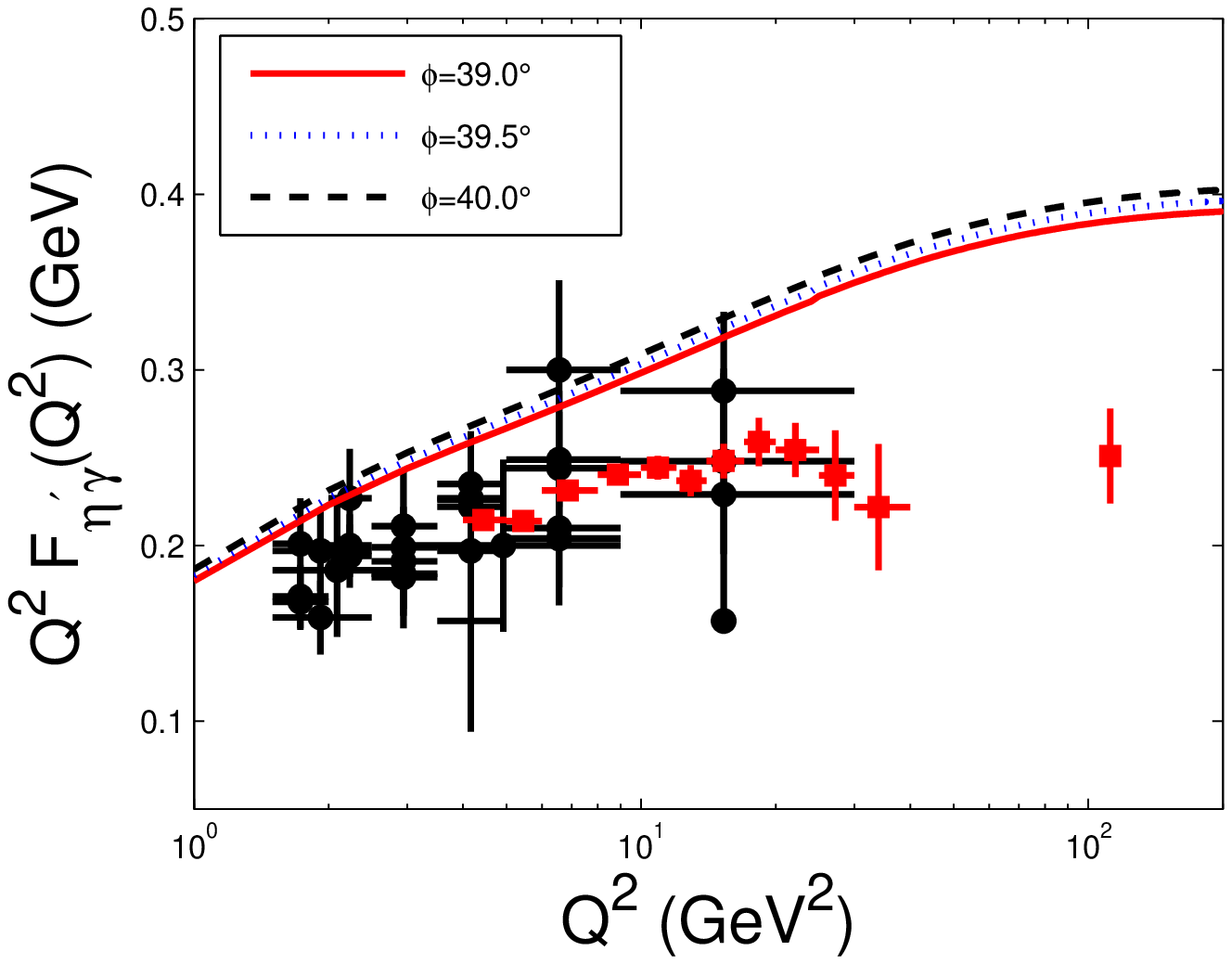}
\caption{$\eta-\gamma$ (Left) and $\eta'-\gamma$ (Right) transition form factors $Q^{2}F_{\eta\gamma}(Q^2)$ and $Q^{2}F_{\eta'\gamma}(Q^2)$ with fixed $B=0.3$ and $m_q=0.30$ GeV and $m_s=0.45$ GeV. The experimental data are taken from Refs.\cite{cleo,babarold,babareta}. } \label{phiuncern}
\end{figure}

The conditions for $\eta-\gamma$ and $\eta'-\gamma$ are somewhat more different. Due to $\eta-\eta'$ mixing, we need to consider these two form factors simultaneously. The curves for $Q^{2}F_{\eta\gamma}(Q^2)$ and $Q^{2}F_{\eta'\gamma}(Q^2)$ for $\phi=39.5^{\circ}\pm 0.5^{\circ}$ are presented in Fig.(\ref{phiuncern}). It is found that $Q^{2} F_{\eta\gamma}(Q^2)$ decreases with the increment of $\phi$, while $Q^{2}F_{\eta'\gamma}(Q^2)$ increases with the increment of $\phi$. By shifting $\phi$ to a smaller value $\sim 38^{\circ}$, one can explain these two form factors within $Q^2<20$ GeV$^2$ consistently as has been pointed out by Ref.\cite{whtheta}. However such a shifting of $\phi$ can not explain the newly BABAR data on $\eta-\gamma$ and $\eta'-\gamma$ for even larger $Q^2>20$ GeV$^2$. So it is hard to fit the gap between the theoretical estimation and the experimental data in the whole $Q^2$ region by a simple variation of $\phi$. Experimentally, $Q^{2}F_{\eta\gamma}(Q^2)$ still increases with the increment of $Q^2$ up to a large value even though its ascending trends is slower than the growth of $Q^{2}F_{\pi\gamma}(Q^2)$, while $Q^{2}F_{\eta'\gamma}(Q^2)$ tends to be a consistent for $Q^2\to\infty$. So some other sources have to be introduced to explain both the $\eta-\gamma$ and $\eta'-\gamma$ form factors in the whole energy region consistently. As shown by Fig.(\ref{eta}), $Q^2 F_{\eta\gamma}(Q^2)$ can agree with the data with $B\sim 0.30$, so we hope the new sources shall have less effects to $Q^2 F_{\eta\gamma}(Q^2)$ than that of $Q^2 F_{\eta'\gamma}(Q^2)$. It has been suggested that a proper intrinsic charm component may have some help to explain the abnormally large production of $\eta'$ \cite{bpi,feldmann,yeh,fc1}. In the following subsection, we shall make a detailed discussion on the possible contributions from the intrinsic charm components.

\subsection{Possible contributions from the intrinsic charm components to $Q^2F_{\eta\gamma}(Q^2)$ and $Q^2F_{\eta'\gamma}(Q^2)$ }

Since the mixing between the $c\bar{c}$ state with $q\bar{q}$-$s\bar{s}$ basis is quite small \cite{feldmann}, we can set
\begin{eqnarray}
\label{qsc1}
F_{\eta\gamma}(Q^2)&=&F_{\eta_q\gamma}(Q^2)\cos\phi
-F_{\eta_s\gamma}(Q^2)\sin\phi +F^{\eta}_{\eta_c\gamma}(Q^2)\\
\label{qsc2}
F_{\eta'\gamma}(Q^2)&=&F_{\eta_q\gamma}(Q^2)\sin\phi
+F_{\eta_s\gamma}(Q^2)\cos\phi +F^{\eta'}_{\eta_c\gamma}(Q^2),
\end{eqnarray}
where $F^{\eta}_{\eta_c\gamma}(Q^2)$ and $F^{\eta'}_{\eta_c\gamma}(Q^2)$ corresponds to the contributions from the intrinsic charm component in $\eta$ and $\eta'$ respectively. Similarly, the WF of the ``intrinsic" charm component $\eta_c=|c\bar{c}\rangle$ can be modeled as
\begin{eqnarray}
\Psi^c_{\eta/\eta'}(x,\mathbf{k}_\perp)&=&A^{c}_{\eta/\eta'}\left(1+B\times C^{3/2}_2(2x-1)\right)
\left[\exp\left(-\frac{{\bf k}_{\perp}^2 +m_c^2}{8{\beta_{c}}^2x(1-x)}\right)\chi^K(m_c,x, {\bf
k}_{\perp})\right],
\end{eqnarray}
where we adopt $\beta_c=\beta_{\pi}|_{m_c}$. The overall factor $A^{c}_{\eta/\eta'}$ is determined by the WF normalization, in which their corresponding decay constants $f^c_\eta$ and $f^c_{\eta'}$ are related by \cite{feldmann}, $\frac{f^c_\eta}{f^c_{\eta'}}=-\tan \left[\phi - \arctan\frac{\sqrt2 f_s} {f_q}\right]$. Here to calculate $F^{\eta}_{\eta_c\gamma}(Q^2)$ and $F^{\eta'}_{\eta_c\gamma}(Q^2)$, the charm quark mass effect should be taken into consideration in the hard part of the amplitude, i.e. the higher helicity states that are proportional to the quark mass shall provide sizable contributions. After integrating over the azimuth angle, a direct calculation shows \cite{whtheta}
\begin{equation}
Q^{2}F_{\eta_{c} \gamma}(Q^2) = \frac{\sqrt{2}}{3\sqrt{3}\pi^2} \int_0^1\frac{d x}{x} \int_0^{\infty}\Psi^{c}_{\eta/\eta'}(x,k_\perp^2) \left[1+\frac{1-z-y^2} {\sqrt{(z+(1-y)^2)(z+(1+y)^2)}}\right] k_\perp
d k_\perp ,
\end{equation}
where $z=\frac{m_c^2}{x^2 Q^2}$ and $y=\frac{k_\perp}{xQ}$.

\begin{figure}
\centering
\includegraphics[width=0.45\textwidth]{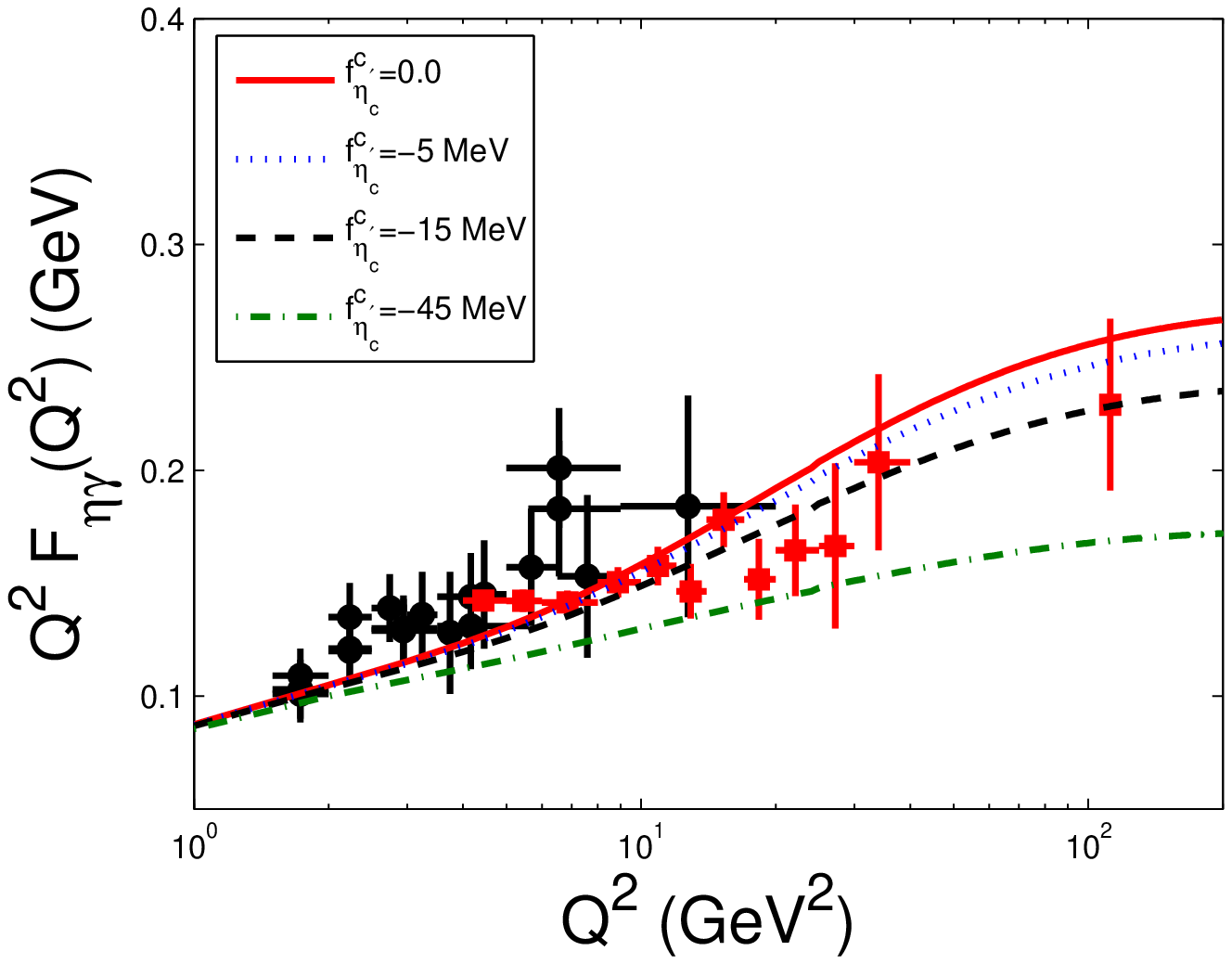}
\hspace{0.1cm}
\includegraphics[width=0.45\textwidth]{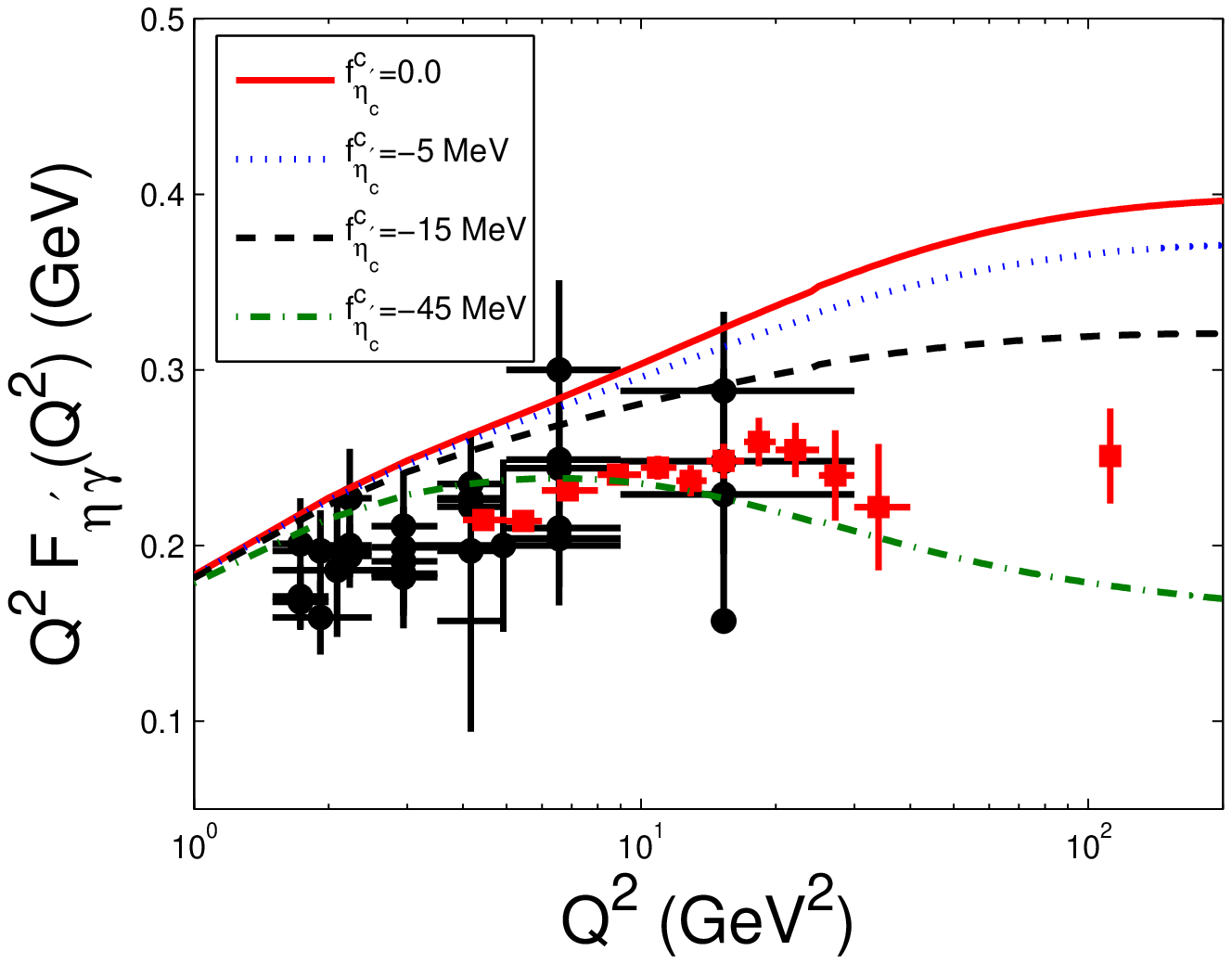}
\caption{$\eta-\gamma$ (Left) and $\eta'-\gamma$ (Right) transition form factors $Q^{2}F_{\eta\gamma}(Q^2)$ and $Q^{2}F_{\eta'\gamma}(Q^2)$ with varying $f^{c}_{\eta'}$, where we take $B=0.30$, $m_q=0.30$ GeV, $m_s=0.45$ GeV and $m_c=1.50$ GeV. The experimental data are taken from Refs.\cite{cleo,babarold,babareta}. } \label{charmA}
\end{figure}

Taking $B=0.30$, $m_q=0.30$ GeV, $m_s=0.45$ GeV, $m_c=1.50$ GeV and $\phi=39.5^{\circ}$, we show how $f_{\eta'}^c$ affects the form factors $Q^2F_{\eta\gamma}(Q^2)$ and $Q^2F_{\eta'\gamma}(Q^2)$. The results are presented in Fig.(\ref{charmA}), where $f^{c}_{\eta'}=0$, $-5$ MeV, $-15$ MeV and $-45$ MeV respectively. These two form factors are slightly affected by the charm component in low $Q^2$ region, while in high $Q^2$ region, the form factors are quite sensitive to $f^{c}_{\eta'}$ and they can be greatly suppressed by possible charm component. One may observe that the experimental data disfavors a larger portion of charm component as $|f_{\eta'}^c| \gtrsim 50$ MeV. And for a more larger $|f^{c}_{\eta'}|$, it shall have more obvious effects to $Q^2 F_{\eta'\gamma}(Q^2)$ than that of $Q^2 F_{\eta\gamma}(Q^2)$, which is what we wanted.

\begin{figure}
\centering
\includegraphics[width=0.45\textwidth]{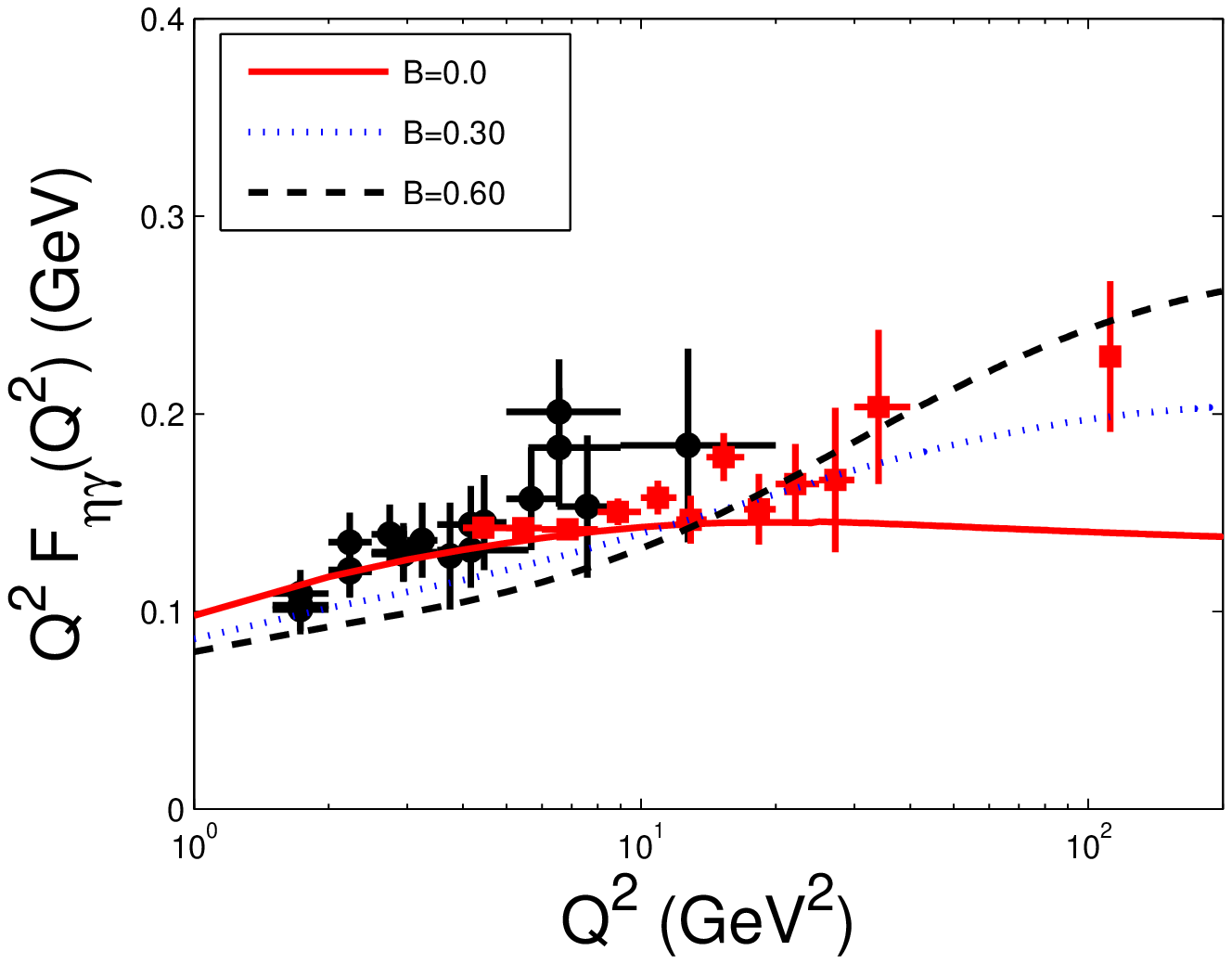}
\hspace{0.1cm}
\includegraphics[width=0.45\textwidth]{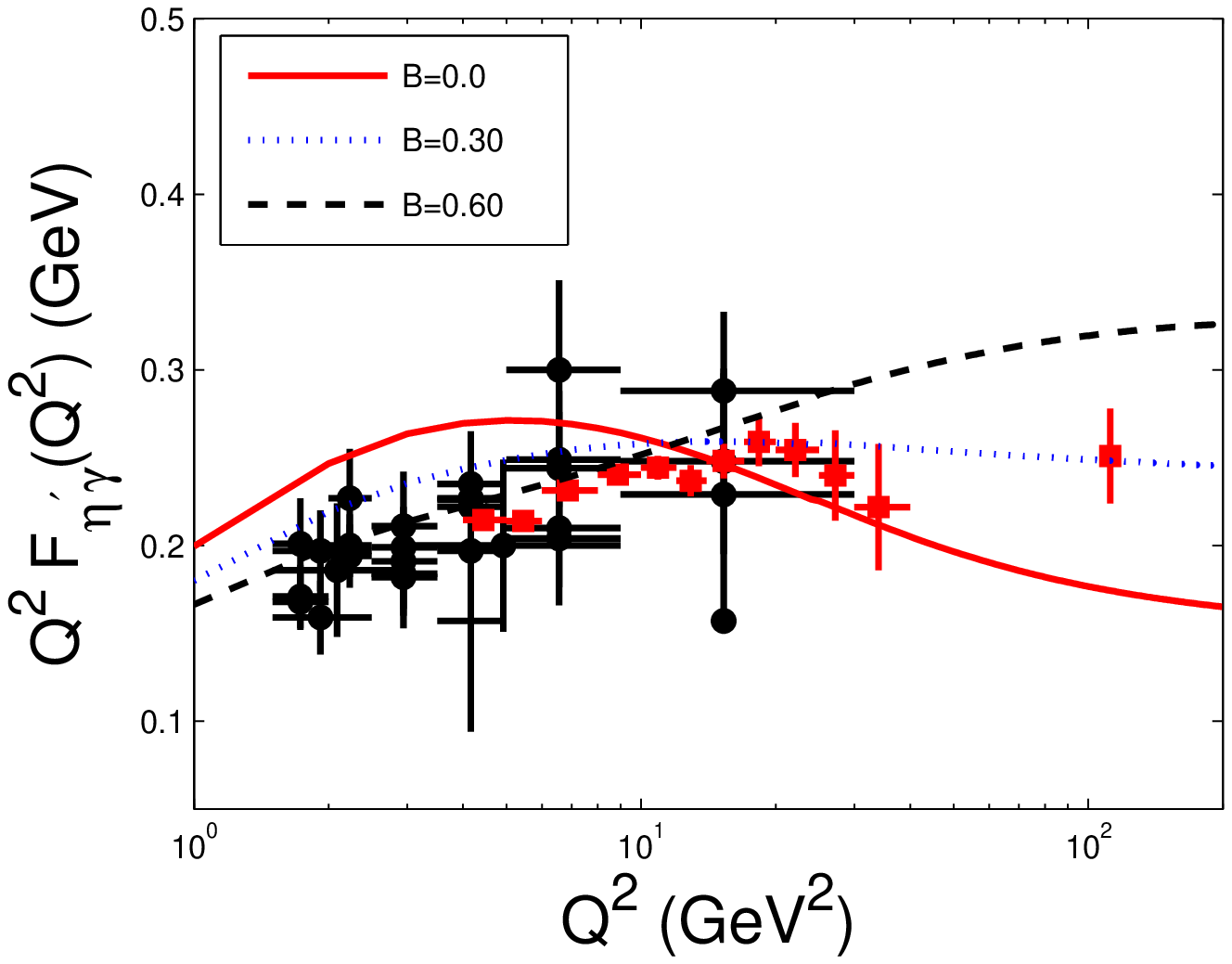}
\caption{$\eta-\gamma$ (Left) and $\eta'-\gamma$ (Right) transition form factors $Q^{2}F_{\eta\gamma}(Q^2)$ and $Q^{2}F_{\eta'\gamma}(Q^2)$ with varying $B$, where $f^{c}_{\eta'}=-30$ MeV, $m_q=0.30$ GeV, $m_s=0.45$ GeV and $m_c=1.50$ GeV. The experimental data are taken from Refs.\cite{cleo,babarold,babareta}. } \label{charmB}
\end{figure}

More over, we present the results for $\eta-\gamma$ and $\eta'-\gamma$ transition form factors with fixed $f^{c}_{\eta'}=-30$ MeV in Fig.(\ref{charmB}), where $B=0.0$, $0.30$ and $0.60$ respectively. It shows that with proper charm component $f^{c}_{\eta'}\sim -30$ MeV and $B\sim 0.30$, the experimental data on $Q^{2}F_{\eta\gamma}(Q^2)$ and $Q^{2}F_{\eta'\gamma}(Q^2)$ can be consistently explained. It is found that $f^{c}_{\eta'}=-30$ MeV is consistent with Ref.\cite{new8,fc1}. Because we still have $|f^{c}_{\eta'}|<<f_{\eta_c} \sim 400$ MeV, according to the mass-matrix-element shown by Ref.\cite{feldmann}, we still have $M_{\eta_c}\simeq m_{cc}$ up to high accuracy. And applying the parameters into the formulas presented in Ref.\cite{feldmann}, it can be found that the mixing between $c\bar{c}$ state with $q\bar{q}-s\bar{s}$ basis is still quite small, i.e. the mixing is around $1\%$. Then our present approximations (\ref{qsc1},\ref{qsc2}) are still reasonable.

\section{Summary}

Light pseudoscalar meson-photon transition form factor provides a good platform to study the leading-twist DA of the light pseudoscalar mesons since it contains only one bound state. In the present paper, we have analyzed three pseudoscalar meson-photon transition form factors consistently by using an uniform WF model suggested by Ref.\cite{wh}. By comparing the estimations with the experimental data on these form factors, it can provide strong constraints on the light pseudoscalar meson DAs. Our results are listed in the following.

(1) According to Eqs.(\ref{fft},\ref{ffv},\ref{ffnv}), all pseudoscalar meson-photon transition form factors $Q^{2}F_{\pi\gamma}(Q^2)$, $Q^{2}F_{\eta_q\gamma}(Q^2)$ and $Q^{2}F_{\eta_s\gamma}(Q^2)$ should have similar behaviors. Since no rapid growth of $Q^{2}F_{\eta\gamma}(Q^2)$ and $Q^{2}F_{\eta'\gamma}(Q^2)$ as that of $Q^{2}F_{\pi\gamma}(Q^2)$ has been found experimentally, these two form factors can be explained by setting $B\sim 0.30$ together with small charm quark component $|f_{\eta'}^c| \sim 30$ MeV. Such a moderate DA with $B\sim 0.30$ for $\eta_q$ and $\eta_s$, which corresponds to the second Gegenbauer DA moment around $0.35$, may also be the pion DA behavior.

(2) We have made a detailed discussion on the form factors' uncertainties caused by the constituent quark masses $m_q$ and $m_s$, the parameter $B$, the mixing angle $\phi$ and the possible intrinsic charm components $f_{\eta}^c$ and $f_{\eta'}^c$. Firstly, the parameter $B$ determines the main behavior of the form factors. By varying $B\in[0.00,0.60]$, one can conveniently obtain the results for $\pi-\gamma$ transition form factor with those obtained in the literature, in which DA behavior varies from AS-like to CZ-like accordingly. A CZ-like DA with $B\sim 0.60$ can explain the data in high $Q^2$ region, however it fails to explain the form factors' lower $Q^2$ behavior. While the AS-like DA with $B\sim 0.00$ can give a better understanding for lower $Q^2$ region, it can not explain the present BABAR data for large $Q^2$ behavior. Secondly, the parameters $m_q$, $m_s$ and $\phi$ can improve the behavior slightly. With the increment of $m_q$ and $m_s$, three form factors $Q^{2}F_{\pi\gamma}(Q^2)$, $Q^{2}F_{\eta\gamma}(Q^2)$ and $Q^{2}F_{\eta'\gamma}(Q^2)$ increase in lower $Q^2$ region and decrease in higher $Q^2$ region. $Q^{2} F_{\eta\gamma}(Q^2)$ decreases with the increment of $\phi$, while $Q^{2}F_{\eta'\gamma}(Q^2)$ increases with the increment of $\phi$. Thirdly, $Q^{2}F_{\eta\gamma}(Q^2)$ and $Q^{2}F_{\eta'\gamma}(Q^2)$ are slightly affected by the charm component in low $Q^2$ region, while in high $Q^2$ region, the form factors are quite sensitive to $f^{c}_{\eta'}$ and they can be greatly suppressed by possible charm component.

(3) It has been found that by adjusting these parameters within their reasonable regions, one can improve the estimations of the form factors to a certain degree but still can not solve the puzzle, especially to explain the behavior of $\pi-\gamma$ form factor within the whole $Q^2$ region consistently. Due to the cancelation between $F_{\eta_{q}\gamma}$ and $F_{\eta_{s}\gamma}$, it is reasonable that $Q^{2}F_{\eta'\gamma}(Q^2)$ tends to be a constant as $Q^2 \to\infty$ \cite{babarold,babareta}. However, it is hard to understand why $Q^{2}F_{\eta\gamma}(Q^2)$ and $Q^{2}F_{\pi\gamma}(Q^2)$ have such a quite different large $Q^2$ behavior. Especially it is hard to explain the rapid growth of $Q^{2}F_{\pi\gamma}(Q^2)$, i.e. probably the logarithmic growth \cite{flatda2,zh}, reported by BABAR collaboration \cite{babar} to be consistent with the previously obtained low $Q^2$ behavior by CELLO and CLEO collaborations \cite{cello,cleo}. Possible charm components $f_{\eta}^c$ and  $f_{\eta'}^c$ can shrink the gap between these two form factors to a certain degree, but it can not be the reason for such a large difference. We hope more experimental data on these form factors in large $Q^2$ region can clarify the present situation. If the BABAR data is confirmed, then there may indeed indicate new physics in these form factors, since it is hard to be explained by the current adopted light-cone pQCD framework.

\hspace{1cm}

{\bf Acknowledgments}: This work was supported in part by the Fundamental Research Funds for the Central Universities under Grant No.CDJZR101000616 and the Program for New Century Excellent Talents in University under Grant No. NCET-10-0882, and by Natural Science Foundation of China under Grant No.10975144, No.10805082 and No.11075225.

\end{document}